\documentclass[preprint,showpacs,preprintnumbers,amsmath,amssymb]{revtex4}
\usepackage{graphicx}% Include figure files
\usepackage{dcolumn}% Align table columns on decimal point
\usepackage{bm}% bold math

\def\be{\begin{equation}}
\def\ee{\end{equation}}
\def\e#1{\label{#1}\end{equation}}
\def\bea{\begin{eqnarray}}
\def\eea{\end{eqnarray}}
\def\ea#1{\label{#1}\end{eqnarray}}
\def\r#1{(\ref{#1})}
\def\bem#1{\begin{mathletters}\label{#1}}
\def\eml{\end{mathletters}}

\def\ket#1{{|#1\rangle}}
\def\bra#1{{\langle#1|}}

\def\4#1{{\boldsymbol{#1}}}
\def\8#1{{\widetilde{#1}}}
\def\mean#1{{\langle{#1}\rangle}}
\def\P{\mathcal{P}}
\def\Q{\mathcal{Q}}
\newcommand{\mc}[1]{\mathcal{#1}}
\def\K{\mc{K}_+}
\def\G{\mc{G}_-}
\def\F{\mc{F}}
\def\t{\Theta}

\def\L{\mc{L}}
\def\rt{\rho_{\text{tot}}}
\def\drt{\dot{\rho}_{\text{tot}}}
\def\lp{\left(}
\def\rp{\right)}

\begin{document}

\title[Universal Dynamical Decoherence Control]{Universal dynamical decoherence control of noisy single-
and multi-qubit systems}

\author{Goren Gordon\footnote{goren.gordon@weizmann.ac.il},
Noam Erez\footnote{nerez@weizmann.ac.il},
Gershon Kurizki\footnote{gershon.kurizki@weizmann.ac.il} }
\address{Department of Chemical Physics,
Weizmann Institute of Science, 76100 Rehovot, Israel}

\begin{abstract}
In this article we develop, step by step, the framework for
universal dynamical control of two-level systems (TLS) or qubits
experiencing amplitude- or phase-noise (AN or PN) due to coupling to a thermal bath. A
comprehensive arsenal of modulation schemes is introduced and
applied to either AN or PN, resulting in completely analogous
formulae for the decoherence rates, thus underscoring the unified
nature of this universal formalism. We then address the extension
of this formalism to multipartite decoherence control, where
symmetries are exploited  to overcome decoherence.
\end{abstract}
%%%\submitto{Journal of Physics B}
\maketitle

\section{Introduction}
\label{sec-intro}
In-depth study of the mechanisms of decoherence and
disentanglement and their prevention in bipartite or multipartite
open systems is an essential prerequisite for applications
involving quantum information processing or communications
\cite{nie00}. The present article is aimed at furthering our
understanding of these formidable issues, which is scanty at best.
It is based on recent progress by our group, as well as others,
towards a unified approach to the dynamical control of decoherence
and disentanglement. This unified approach culminates in universal
formulae allowing design of the required control fields.

The topic of multipartite decoherence has been well-investigated
in two limits. One of these is relaxation toward steady-state of
one-body coherence of spins, atoms, excitons, quantum dots, etc.
that are in contact with a much larger reservoir. The other is the
collective decoherence of a small (\emph{localized}) two-body or
many-body system, which typically occurs more rapidly than
one-body decoherence \cite{coh92,scu97}.

By contrast, more general problems of decay of \emph{non-local}
mutual entanglement of two or more small systems are less well
understood. This decoherence process may occur on a time scale
much shorter than the time for either body to undergo local
decoherence, but much larger than the time each takes to become
disentangled from its environment. The disentanglement of
individual particles from their environment is dynamically
controlled by interactions on non-Markovian time-scales, as
discussed below \cite{aku05}. Their disentanglement from each
other, however, may be purely Markovian \cite{ban04a,yu04,yu06},
in which case the present non-Markovian approach to dynamical
control/prevention is insufficient.

\subsection{Dynamical control of decay and decoherence on
non-Markovian time scales}
\label{ch-1-sec-1}
Quantum-state decay to a continuum or changes in its population
via coupling to a thermal bath is known as amplitude noise (AN).
It characterizes decoherence processes in many quantum systems,
e.g., spontaneous emission of photons by excited atoms
\cite{coh92}, vibrational and collisional relaxation of trapped
ions \cite{sac00} and the relaxation of current-biased Josephson
junctions \cite{cla88}. Another source of decoherence in the same
systems is proper dephasing or phase noise (PN) \cite{scu97},
which does not affect the populations of quantum states, but
randomizes their energies or phases.

A thoroughly studied approach to suppression of decoherence is the
``dynamical decoupling'' of the system from the bath
\cite{aga99,aga01a,alicki2004oss,vio98,shi04,vit01,fac01,fac04,zan03}.
In particular, ``bang-bang'' (BB) pulses have been proposed for
\emph{stroboscopic} suppression of proper dephasing: $\pi$-phase
flips of the coupling via strong and sufficiently fast resonant
pulses applied to the system \cite{vio98,shi04,vit01}. The
identification of a decoherence-free subspace (DFS), wherein
symmetrically degenerate states are decoupled from the bath,
constitutes a complementary approach
\cite{zan03,zan97,lid98,wu02,kof96}.

Our group has purported to substantially expand the arsenal of
decay and decoherence control. We have presented a {\em universal
form of the decay rate} of unstable states into {\em any}
reservoir (continuum), dynamically modified by perturbations with
arbitrary time dependence, focusing on non-Markovian time-scales
\cite{kof96,kof00,kof01,kof04,kof05}. An analogous form has been
obtained by us for the dynamically modified rate of proper
dephasing \cite{kof04,kof05,kof01a}. Our unified, optimized
approach reduces to the BB method in the particular case of proper
dephasing or decay via coupling to {\em spectrally symmetric}
(e.g., Lorentzian or Gaussian) noise baths with limited spectral
width (see below). The type of phase modulation advocated for the
suppression of coupling to {\em asymmetric} baths (e.g., phonon or
photon baths with frequency cutoff \cite{pel04}) is, however,
drastically different from the BB method. Other situations to
which our approach applies, but not the BB method, include {\em
amplitude modulation} of the coupling to the continuum, as in the
case of decay from quasibound states of a periodically tilted
washboard potential \cite{kof01}: such modulation has been
experimentally shown \cite{fis01} to give rise to either slowdown
of the decay (Zeno-like behavior) or its speedup (anti-Zeno-like
behavior), depending on the modulation rate.

The theory has been generalized by us to finite temperatures and
to qubits driven by an {\em arbitrary} time-dependent field, which
may cause the failure of the rotating-wave approximation
\cite{kof04}. It has also been extended to the analysis of {\em
multi-level systems}, where quantum interference between the
levels may either inhibit or accelerate the decay \cite{gor06a}.

Our general approach \cite{kof01} to dynamical control of states
coupled to an arbitrary ``bath'' or continuum has reaffirmed the
intuitive anticipation that, in order to suppress their decay, we
must modulate the system-bath coupling at a rate exceeding the
spectral interval over which the coupling is significant. Yet our
analysis can serve as a general recipe for {\em optimized} design
of the modulation aimed at an effective use of the fields for
decay and decoherence suppression or enhancement. The latter is
useful for the control of chemical reactions \cite{pre00}.

\subsection{Control of symmetry-breaking multipartite decoherence}
\label{ch-1-sec-2}
Symmetry is a powerful means of protecting entangled quantum
states against decoherence, since it allows the existence of a
decoherence-free subspace or a decoherence-free subsystem
\cite{cla88,scu97,aga99,aga01,aga01a,vio98,shi04,vit01,fac01,fac04,zan03,zan97,lid98,wu02,kof96,vio99,vio00,aku05}.
In multipartite systems, this requires that all particles be
perturbed by the {\em same} environment. In keeping with this
requirement, quantum communication protocols based on entangled
two-photon states have been studied under {\em collective}
depolarization conditions, namely, {\em identical} random
fluctuations of the polarization for both photons \cite{ban04}.
Entangled qubits that reside at the same site or at equivalent
sites of the system, e.g. atoms in optical lattices, have likewise
been assumed to undergo identical decoherence \cite{aku05}.

Locally-decohering entangled states of two or more particles, such
that each particle travels along a different channel or is stored
at a different site in the system, may break the state symmetry. A
possible consequence of this symmetry breaking is the abrupt
``death'' of the entanglement \cite{ban04a,yu04,yu06}. Such
systems, composed of particles undergoing individual or ``local''
decoherence, do not possess a natural DFS and thus present more
challenging problems insofar as decoherence effects are concerned
\cite{lis02}.

Our group has recently addressed these challenges by developing a
generalized treatment of multipartite entangled states (MES)
decaying into zero-temperature baths and subject to {\em
arbitrary} external perturbations whose role is to provide {\em
dynamical protection} from decay and decoherence \cite{gor06a,
gor06b}. Our treatment applies {\em to any difference} between the
couplings of individual particles to the baths. It does not assume
the perturbations to be stroboscopic, i.e. strong or fast enough,
but rather to act concurrently with the particle-bath
interactions. Our main results are to show that by applying {\em
local} (selective) perturbations to multilevel particles, i.e. by
{\em addressing each level and each particle individually}, one
can create a decoherence-free system of many entangled qubits.
Alternatively, one may reduce the problem of locally decohering
MES to that of a single decohering particle, whose dynamical
control has been thoroughly investigated
\cite{vio98,shi04,kof01,kof04,kof05}. On the other hand, the
combined effect of dephasing and relaxation (phase and amplitude
noises) on MES and its control, which constitute a much more
formidable problem, have not yet been studied by us.

\subsection{Outline}
\label{Subsec-layout} In this article we develop, step by step, the framework for
universal dynamical control by modulating fields of two-level
systems or qubits, aimed at suppressing or preventing their noise,
decoherence or relaxation in the presence of a thermal bath. To
this end, a comprehensive treatment is developed in
Sec.~\ref{Sec-ME} in a more complete and transparent fashion than
its brief sketch in Ref.\cite{kof04}.  Its crux is the derivation
of a more general master equation (ME) than in previous treatments
of a multilevel, multipartite system, weakly coupled to an
arbitrary bath and subject to arbitrary temporal driving or
modulation. The present ME, derived by the Nakajima-Zwanzig
technique \cite{nakajima1958qtt,zwanzig1960emt}, is more general
than the ones obtained previously in that it does not invoke the
rotating wave approximation and therefore applies at arbitrarily
short times or for arbitrarily fast modulations.

Remarkably, when our general ME is applied to either AN or PN in
Sec.~\ref{Sec-Bloch}, the resulting dynamically-controlled
relaxation or decoherence rates obey \emph{analogous formulae}
provided the the corresponding density-matrix (generalized Bloch)
equations are written in the appropriate basis. This underscores
the universality of our treatment. The choice of an appropriate
time-dependent basis allows here to simplify the AN treatment of
Ref. \cite{kof04}. More importantly, it allows us to present a PN
treatment that does not describe noise phenomenologically as in
Ref. \cite{kof04}, but rather dynamically, starting from the
ubiquitous spin-boson Hamiltonian.

We then discuss in Sec.~\ref{Sec-modulation}, more comprehensively
than in previous treatments, the possible modulation arsenal for
either AN or PN control. The present formalism is applicable in a
natural and straightforward manner to multipartite and/or
multilevel systems \cite{gor06b}. It allows us to focus in
Sec.~\ref{Sec-multi} on the ability of symmetries to overcome
multipartite decoherence\cite{zan03,zan97,lid98,wu02,vio00}.  Our
conclusions are presented in Sec.~\ref{Sec-conc}.

%%%%%%%%%%%%%%%%%%%%%%%%%%%%++++++++++++++++++++++++++++++++++++++%%%%%%%%%%%%%%%%%%%%%%%%%%%%%%%%%%%%%%%%%%%%%%%%%%%%
\section{Master equation (ME) for dynamically controlled systems coupled to thermal baths}
\label{Sec-ME}
\subsection{Derivation of the reduced density matrix ME by the Nakajima-Zwanzig method}

We shall consider the most general unitary evolution of a system
coupled to a thermal reservoir, governed by the Liouville operator
equation (we shall take $\hbar=1$ throughout the rest of the
paper):

\bea
\drt(t) &=&-i [H(t),\rho_{tot}(t)] \equiv -i \L(t) \rho_{tot}(t), \label{Leq}
\eea
\begin{subequations}
\bea
H(t) &=& H_0(t)+H_I(t), \\
H_0(t)&\equiv& H_S(t) +H_B,
\eea
\end{subequations}
\bea
\L(t)&=& \L_S(t)+\L_B+\L_I(t)
\eea where $\rho_{tot}$ is the density matrix of the system+reservoir and
$H_S,~H_B,~H_I$ are the Hamiltonians of the system, bath and their
interaction, respectively. As usual, $\L$ denotes the Liouville
operator, which acts linearly on \emph{operators} on our Hilbert
space. We shall use the notation $H_0(t)$ for the ``unperturbed''
Hamiltonian, assuming weak system-bath coupling, and $\L_0(t)$ for
its associated Liouville operator.

We seek a master equation for the reduced density matrix of the
system alone $\rho\equiv Tr_B \rho_{tot}$, allowing for
\emph{arbitrary} time dependence of $H(t)$ \emph{without}
resorting to the rotating-wave approximation \cite{coh92,scu97}.
This can be accomplished by the Nakajima-Zwanzig
\cite{nakajima1958qtt,zwanzig1960emt,prigogine1962nes}
projection-operator technique (see also
\cite{breuer2002toq,yan2005qmd}). Let us define $\rho_B\equiv
Z^{-1}e^{-H_B/k_B T}$ ($T$ being the reservoir temperature, $Z$
normalization to unit trace), the projection operator
$\P(\cdot)\equiv \text{Tr}_B(\cdot)\otimes \rho_B$ (satisfying
$\P^2=\P$), and the complementary (projection) operator $\Q \equiv
1-\P$.

%We shall follow the derivation of Hashitsume et al.[] of their ``time convolutionless''(TCL) form of the Master equation, and its form under the Born approximation.
%In this subsection, we shall assume the Hamiltonian is time independent.

In terms of these definitions, Eq.\eqref{Leq} is equivalent to:

\begin{subequations}
\bea
\label{e2a}\P \drt (t) &=& -i\P \L(t) \P \rt(t) -i \P \L(t) \Q \rt (t) \\ \Q
\drt (t) &=& -i\Q \L(t) \P \rt(t) -i \Q \L(t) \Q \rt (t)
\label{e2b}
\eea
\end{subequations}
Equation \eqref{e2b} is then formally integrated to give:

\be
\Q \rt(t) = -i\int_0^t \K(t,\tau) \Q \L(\tau) \P \rt(\tau)d\tau
+\K(t,0)\Q \rt(0) \label{e3}
\ee where

\be
\K(t,\tau) = \text{T}_+e^{-i\Q\int_\tau^t \L(s)ds},
\ee $\text{T}_+$ denoting time-ordering.

If this expression is then plugged into Eq.\eqref{e2a} we get a non-Markovian ME for $P\rt$:

\be
\P \drt(t) = -i\P \L(t)\P \rt(t) - \int_0^t \P \L(t) \K(t,\tau) \Q \L(\tau)\P \rt (\tau) d\tau -i\P \L \K(t,0)\Q \rt(0)
\ee which yields a ME for $\rho$ after tracing out the bath.

Rather than apply a perturbative treatment directly to this
equation, it is useful to first transform it to a
``time-convolutionless'' form (TCL)
\cite{hashitsumae1977qme,shibata1977gsl,chaturvedi1979tcp,shibata1980efn}.
In this form, the memory effect (the presence of $\rt(\tau)$ in
the integrand) is transferred to the integration kernel and
$\rho(t)$ is taken out of the integral. Only then, is the
perturbative expansion (in $\L_I$) applied.

Formally,

\be
\rt(\tau) = \G(t,\tau) \rt(t)
\ee where, writing $\text{T}_-$ for anti-chronological ordering:

\be
\G(t,\tau) \equiv \text{T}_-e^{+i\int_\tau^t \L(s) ds}.
\ee Substituting this expression for $\rt(\tau)$ into Eq.\eqref{e3}, one obtains:

\begin{subequations}

\be
\Q \rt(t) = - \int_0^t \K(t,\tau)i\Q \L(\tau)\P \G(t,\tau)d\tau \lp\P+\Q \rp \rt (t) + \K(t,0)\Q \rt(0)
\ee Collecting all $\Q \rt$ terms on the left, we obtain:

\bea
&\F(t)&\Q \rt (t) = \left\{ 1-\F(t)\right\} \P \rt (t) +\K(t,0)\Q \rt (0) \\
&\F(t)& = 1 + \int_0^t \K(t,\tau) i \Q \L(\tau) \P \G(t,\tau)d \tau \equiv 1+ \Sigma(t) \label{e10c}
\eea Assuming $\F(t)$ can be inverted (which is expected to hold for short times in the weak coupling limit),
and writing $\t(t) = \F(t)^{-1}$, one obtains the equation:

\be
\Q \rt (t) = \left\{ \t(t)-1\right\} \P \rt (t) +\t(t) \K(t,0) \Q \rt (0).
\ee \end{subequations} Finally, plugging this expression for $\Q \rt$ into Eq.\eqref{e2a}, the formal TCL ME is obtained:

\begin{subequations}
\be
\P \drt (t) = -i \P \L(t)\P \rt(t) -i \P \L\left\{ \t(t)-1 \right\} \P\rt(t) - i\P \L \t(t)\K(t,0) \Q \rt(0). \label{e11a}
\ee If the initial condition is such that $\Q\rt(0)=0$, so that the last term vanishes (as will indeed be our assumption below), then all
memory effects are contained in $\t(t)$. In what follows, \emph{we shall always assume that}
\be
\rt(0)=\rho_S(0)\otimes\rho_B,
\ee so that this condition is fulfilled.
\end{subequations}

The operators $\L_S$ and $\L_B$ both commute with $\P$ and $\Q$, and $\P \L_B =0$. This implies $\P \L \Q = \P \L_I \Q$ (and note that
$\t - 1 = \Q (\t -1)$). With the notation $\langle \cdot \rangle_B \equiv \text{Tr}_B\lp\cdot \rho_B \rp$, the ME for the reduced
density matrix of the system can be written in the form:

\begin{subequations}

\be
\dot{\rho}(t) = -i \left[\L_S +\langle \L_I \rangle_B \right]\rho(t) - \Xi(t)\rho(t) \label{ME}
\ee

\be
\Xi (t) = \langle i\L_I\left\{\t(t)-1\right\} \rangle_B \label{Xi}
\ee

\end{subequations}

\subsection{Born approximation}

%We assume the initial condition that $\rt(0)=\rho_S(0)\rho_B$, so that the condition $\Q\rt(0)=0$ is fulfilled.

At this point, it is expedient to follow the perturbational method of \cite{emch1968nsm,PhysRev.178.2025}.
We begin by noting that (cf. Eqs.\eqref{e10c},\eqref{Xi}):

\be
\Xi (t) = -i \langle \L_I\left\{\Sigma(t) + \mc{O}(\Sigma^2)\right\} \rangle_B ;~~\Sigma(t)=\mc{O}(\L_I).
\ee Therefore, to expand Eq.\eqref{Xi} to second order in $\L_I$, we need to evaluate $\K$ and $G$ only to 0th order!
The operator $\K(t,\tau)$ can be factored:

\begin{subequations}

\be
%\K(t,\tau) = \underset{\equiv \mc{V}_0(t,\tau)}{\underbrace{\left\{ \text{T}_+ e^{-i\Q\int_\tau^t L_0(s) ds}\right\} } }
%\left\{ \text{T}_+ e^{-i\Q\int_\tau^t \mc{V}_0(s,\tau)^{-1}\L_I(s)\mc{V}_0(s,\tau)   ds} \right\}
\K(t,\tau) = \mc{V}_0(t,\tau) \text{T}_+ e^{-i\Q\int_\tau^t \mc{V}_0(s,\tau)^{-1}\L_I(s)\mc{V}_0(s,\tau)   ds}
\ee where

\be
\mc{V}_0(t,\tau)=\text{T}_+ e^{-i\Q\int_\tau^t \L_0(s) ds}.
\ee \emph{To zeroth order in $\L_I$, $\K$ is just $\mc{V}_0$}. Furthermore, it is important to note that

\end{subequations}

\begin{subequations}

\be
\mc{V}_0 \Q = \Q\mc{U}_0\Q = \Q\mc{U}_0 \label{e14a}
\ee where

\be
\mc{U}_0(t,\tau) = \text{T}_+e^{-i\int_\tau^t \L_0(s)ds},
\ee and in Eq.\eqref{e14a} we have used the fact that $\mc{U}_0$ commutes with $\Q$.
\end{subequations}
Similarly,

\be
\G(t,\tau) = \lp 1+ \mc{O}(\L_I) \rp \text{T}_- e^{+i \int_\tau^t \L_0(s) ds} = \mc{U}_0(t,\tau)^{-1} + \mc{O}(\L_I).
\ee
We have for $\Sigma(t)$ in the Born approximation:

\be
\Sigma(t) = i\int_0^t \mc{V}_0(t,\tau)\Q \L_I(\tau) \P \mc{U}_0(t,\tau)^{-1}d\tau.
\ee
Finally, after making use of Eq.\eqref{e14a}, we get the ME for $\rho$ in the Born approximation, which implies the
neglect of the back-effect of the system on the bath, consistently with the weak-coupling assumption:

\be
\dot{\rho}(t) = -i\lp \L_S(t)+\langle \L_I\rangle_B\rp \rho(t) - \int_0^t\langle \L_I(t)\Q
\mc{U}_0(t,\tau)\L_I(\tau)\P \mc{U}_0(t,\tau)^{-1}\rangle_Bd\tau \rho(t)  \label{MEBorn}
\ee for $\Q\rt(0)=0$.

%%%%%%%%%%%%%%%%%%%%%%%%%%%%%%%%%%%%%%%%%%%%%%%%%%%%%%%%%%%%%%%%%%%%%%%%%%%%

\subsection{Explicit equations for factorizable interaction Hamiltonians}

We now wish to write the ME explicitly for time-dependent
Hamiltonians of the following form \cite{kof04}:

\begin{subequations}

\bea \label{general}
H(t) &=& H_S(t)+H_B+H_I(t), \\
H_I(t) &=& S(t) B;
\eea
\end{subequations} where $H_S$ and $H_B$ are the system and bath Hamiltonians, respectively; and $H_I$,
the interaction Hamiltonian, is the product of operators $S,B$
which act on the system (resp. bath) alone. We assume $\langle
\L_I \rangle_B =0$, by virtue of $\langle B \rangle_B =0$. This
also implies $\P \L_I\P =0$. Equation \eqref{MEBorn} now
simplifies to:

\be
\dot\rho(t) = -i\L_S(t)\rho(t) - \int_0^t d\tau \langle \L_I(t) \mathcal{U}_0(t,\tau) \L_I(\tau)
\P \mathcal{U}_0(t,\tau)^{-1} \rangle_B  \rho(t). \label{e20}
\ee

Let us now write out the action of the operator $\mc{U}_0$ in terms of the unitary evolution
operators of the system and bath:

\begin{subequations}

\be
\mc{U}_0(t,\tau)A = U_S(t,\tau) U_B(t-\tau)A U_B(\tau-t)  U_S(t,\tau)^\dagger \\
\ee

\be
\label{U-s-def}
U_S(t,\tau) \equiv \text{T}_+e^{-i\int_\tau^t H_S(t')dt'}
\ee

\be
U_B(t) \equiv e^{-iH_Bt}.
\ee

\end{subequations}

The integrand of the second term in Eq.\eqref{e20}, can now be written explicitly as:

\bea
I(t,\tau)&=&\text{Tr}_B \left[ S(t)B, \mc{U}_0(t,\tau)
\left[S(\tau)B,\mc{U}_0(t,\tau)^{-1}\rho(t)\rho_B
\right]\right] \nonumber \\
&=& \text{Tr}_B  \left[ S(t)B, \left[\tilde{S}(t,\tau)\tilde{B}(t-\tau),\rho(t) \rho_B \right]\right]
\eea where $\tilde{S}$  and $\tilde{B}$ are defined as:

\bea
\label{S-tilde-def}
\tilde{S}(t,\tau)&\equiv& U_S(t,\tau)S(\tau)U_S(t,\tau)^\dagger, \\
\tilde{B}(\tau)&\equiv& U_B(\tau)BU_B(-\tau).
\eea
Using the commutativity of $S$ and $B$, and of $\rho_B$ and $H_B$,
as well as the cyclic property of the trace, this gives after some
rearrangement:

\be
I(t,\tau) = \langle B\tilde{B}(t-\tau)\rangle_B
\left[S(t),\tilde{S}(t,\tau)\rho(t)\right] + H.c.
\ee

Finally, defining the correlation function for the bath,
\begin{subequations}
\be
\Phi_T(t) = \langle B \tilde{B}(t) \rangle_B,
\ee we obtain the ME for $\rho$ in the Born approximation:

\be
\label{gen-ME}
\dot{\rho}(t) = -i\left[H_S,\rho(t)\right]+\int_0^t d\tau \left\{
\Phi_T(t-\tau) \left[\tilde{S}(t,\tau)\rho(t),S(t)\right] +H.c.
\right\}.
\ee
\end{subequations}
%%%%%%%%%%%%***********************************************************************************8

\section{Generalized Bloch equations}
\label{Sec-Bloch}
Having derived the master equation, we focus on two regimes: a
two-level system coupled to either an amplitude- or phase-noise
(AN or PN) thermal bath. The bath Hamiltonian (in either regime)
will be explicitly taken to consist of harmonic oscillators and be
linearly coupled to the system (generalizations to other baths and
couplings are obvious):
\bea
&H_B=\sum_\lambda\omega_\lambda a_\lambda^\dagger a_\lambda\\
&B=\sum_\lambda(\kappa_\lambda
a_\lambda+\kappa_\lambda^*a_\lambda^\dagger).
\eea
Here $a_\lambda,a_\lambda^\dagger$ are the annihilation and
creation operators of mode $\lambda$, respectively, and
$\kappa_\lambda$ is the coupling amplitude to mode $\lambda$.

We use different modulation schemes for each regime, namely,
dynamical {\em off-resonant} fields for the AN regime and
time-dependent {\em resonant} fields for the PN regime. We derive
the generalized Bloch equations for the two cases.

\subsection{Two-level system coupled to a thermal amplitude-noise
bath}
\label{Subsec-amplitude-bloch}
We first consider the AN regime of a two-level system coupled to a
thermal bath. We will use off-resonant dynamic modulations,
resulting in AC-Stark shifts. The Hamiltonian
(Eq.~\eqref{general}) then assumes the following form:
\bea
\label{sigma-x-1}
&H_S(t)=(\omega_a+\delta_a(t))\ket{e}\bra{e}\\
\label{sigma-x-2}
&S(t)=\tilde\epsilon(t)\sigma_x
\eea
where $\delta_a(t)$ is the dynamical AC-Stark shifts,
$\tilde{\epsilon}(t)$ is the time-dependent modulation of the
interaction strength, and the Pauli matrix
$\sigma_x=\ket{e}\bra{g}+\ket{g}\bra{e}$.

We derive the Bloch equations for the explicit case discussed
above. Inserting Eqs.~\eqref{sigma-x-1}-\eqref{sigma-x-2} into
Eq.~\eqref{U-s-def} and Eq.~\eqref{S-tilde-def}, we get:
\bea
U_S(t,\tau)&=&e^{-i\omega_a(t-\tau)-i\int_\tau^t dt_1 \delta_a(t_1)}\ket{e}\bra{e}+\ket{g}\bra{g}\\
\tilde{S}(t,\tau)&=&e^{-i\omega_a(t-\tau)-i\int_\tau^t dt_1
\delta_a(t_1)}\tilde{\epsilon}(\tau)\ket{e}\bra{g}+H.c.
\eea
Plugging this into the ME \eqref{gen-ME}, we arrive at the
following modified Bloch equations:
\bea
\label{AN-Bloch-1}
\dot\rho_{ee}=-\dot\rho_{gg}&=&-R_e(t)\rho_{ee}+R_g(t)\rho_{gg} \\
\label{AN-Bloch-2}
\dot\rho_{eg}=\dot\rho_{ge}^*&=&-\left\{(R(t)+i\Delta_a(t))+i[\omega_a+\delta_a(t)]\right\}\rho_{eg}
\nonumber\\&&+\left\{R(t)-i\Delta_a(t)\right\}\rho_{ge},
\eea where
\bea
&&R(t)=[R_{e}(t)+R_{g}(t)]/2\\
&&\Delta_a(t)=\Delta_{e}(t)-\Delta_{g}(t)\\
\label{R-eg}
&&R_{e(g)}(t)/2+i\Delta_{e(g)}(t)=\int_0^tdt'\Phi_T(t-t')K_{e(g)}(t,t')e^{\pm i\omega_a(t-t')}\\
&&K_e(t,t')=K_g^*(t,t')=\epsilon(t)\epsilon^*(t')\\
&&\epsilon(t)=\tilde{\epsilon}(t)e^{i\int_0^tdt_1\delta_a(t_1)}.
\eea
$R_{e(g)}(t)$ is the modified downward (upward) transition rate of
the excited (ground) state to the ground (excited) state. Their
half-rate contributes to the decoherence rate, and $\Delta_a(t)$
is the resonance (transition frequency) shift in energy due to the
modified coupling to the bath.

\subsection{Two-level system coupled to thermal phase-noise bath}
\label{Subsec-phase-bloch}
Next, we consider the PN regime of a two-level system coupled to
thermal bath, where we will use near-resonant fields with
time-varying amplitude as our control. The Hamiltonians
(Eq.~\eqref{general}) then assume the following forms:
\bea
\label{sigma-z}
&H_S(t)=\omega_a\ket{e}\bra{e}+V(t)\sigma_x\\
&S(t)=\tilde\epsilon(t)\sigma_z
\eea
where $V(t)=V_0(t)e^{-i\omega_at}+c.c$ is the time-dependent
resonant field, with real envelope $V_0(t)$, $\tilde{\epsilon}(t)$
is the time-dependent modulation of the interaction strength,
$\sigma_z=\ket{e}\bra{e}-\ket{g}\bra{g}$.

Since we are interested in dephasing, phases due to the
(unperturbed) energy difference between the levels are immaterial.
We eliminate this dependence by moving to the rotating frame. To
avoid the need to time-order the propagator of the system
Hamiltonian we tilt the rotating frame to the time-dependent
basis:
\be
\ket{\uparrow}=\frac{1}{\sqrt{2}}\left(
e^{-i\omega_at}\ket{e}+\ket{g}\right)\quad\ket{\downarrow}=\frac{1}{\sqrt{2}}\left(
e^{-i\omega_at}\ket{e}-\ket{g}\right)
\ee
In this frame, the system and bath Hamiltonians become:
\bea
\label{sigma-hat-z-1}
&\hat{H}_S(t)=\frac{V_0(t)}{2}\hat\sigma_z\\
\label{sigma-hat-z-2}
&\hat{S}(t)=\tilde\epsilon(t)\hat\sigma_x
\eea
where $\hat{}$ denotes the rotated and tilted frame,
$\hat\sigma_z=\ket{\uparrow}\bra{\uparrow}-\ket{\downarrow}\bra{\downarrow}$
and
$\hat\sigma_x=\ket{\uparrow}\bra{\downarrow}+\ket{\downarrow}\bra{\uparrow}$.

We can now derive the Bloch equations for the PN regime discussed
above, and demonstrate their analogy to their AN counterparts \eqref{AN-Bloch-1},\eqref{AN-Bloch-2}.
To this end we insert Eqs.~\eqref{sigma-hat-z-1}-\eqref{sigma-hat-z-2}
into Eq.~\eqref{U-s-def} and Eq.~\eqref{S-tilde-def}, to get:
\bea
U_S(t,\tau)&=&e^{-i\int_\tau^t dt_1 V_0(t_1)/2}\ket{\uparrow}\bra{\uparrow}+e^{i\int_\tau^t dt_1 V_0(t_1)/2}\ket{\downarrow}\bra{\downarrow}\\
\tilde{S}(t,\tau)&=&e^{-i\int_\tau^t dt_1
V_0(t_1)}\tilde{\epsilon}(t)\ket{\uparrow}\bra{\downarrow}+H.c.
\eea
Plugging this into the ME \eqref{gen-ME}, we arrive at the
following modified Bloch equations:
\bea
\dot\rho_{\uparrow\uparrow}&=&-\dot\rho_{\downarrow\downarrow}=-R_\uparrow(t)\rho_{\uparrow\uparrow}+R_\downarrow(t)\rho_{\downarrow\downarrow} \\
\dot\rho_{\uparrow\downarrow}&=&\dot\rho_{\downarrow\uparrow}^*=-\left\{(R(t)+i\Delta_a(t))+iV_0(t)/2\right\}\rho_{\uparrow\downarrow}
+\left\{R(t)-i\Delta_a(t)\right\}\rho_{\downarrow\uparrow},
\eea where
\bea
&&R(t)=[R_{\uparrow}(t)+R_{\downarrow}(t)]/2\\
&&\Delta_a(t)=\Delta_{\uparrow}(t)-\Delta_{\downarrow}(t)\\
\label{R-arrows}
&&R_{\uparrow(\downarrow)}(t)/2+i\Delta_{\uparrow(\downarrow)}(t)=\int_0^tdt'\Phi_T(t-t')K_{\uparrow(\downarrow)}(t,t')\\
&&K_\uparrow(t,t')=K_\downarrow^*(t,t')=\epsilon(t)\epsilon^*(t')\\
\label{PN-epsilon}
&&\epsilon(t)=\tilde{\epsilon}(t)e^{i\int_0^tdt_1V_0(t_1)}.
\eea

As can be clearly seen, these modified Bloch equations are
completely analogous to their AN counterparts,
Eqs.~\eqref{AN-Bloch-1},\eqref{AN-Bloch-2}, provided we change the
basis as follows:
\be
e\leftrightarrow\uparrow\quad g\leftrightarrow\downarrow.
\ee
Despite their analogy, Eqs.~\eqref{R-eg} and \eqref{R-arrows} are \emph{not
identical}, due to the use of the rotating frame in the PN
case. Nevertheless, this analogy underscores the universality of our
approach.

\subsection{Spectral domain representation}
For both AN and PN regime, one can have a more insightful
representation of the modified rates by transforming them to the
frequency domain. For the long time limits (see
Sec.~\ref{Sec-multi}), one arrives at the form
\bea
\label{53}
&&R_{e(g)}(t)=2\pi\int_{-\infty}^\infty d\omega
G_T(\pm(\omega_a+\omega))F_t(\omega)\\
&&R_{\uparrow(\downarrow)}(t)=2\pi\int_{-\infty}^\infty d\omega
G_T(\pm\omega)F_t(\omega)
\eea
where the difference is due to the fact that we used the rotating
and tilted frame in the PN regime. Here $G_T(\omega)$ is the
temperature-dependent bath coupling spectrum given by
\be
G_T(\omega)=(2\pi)^{-1}\int_{-\infty}^\infty dt\Phi_T(t)e^{i\omega
t}.
\ee
Introducing the control-field fluence $Q(t)$, the spectral
modulation $F_t(\omega)$ can be normalized to unity:
\bea
\label{TLS-Fluence}
&Q(t)=\int_0^t
d\tau |\epsilon(\tau)|^2,\\
&F_t(\omega)=\frac{|\epsilon_t(\omega)|^2}{Q_(t)}, \ea{TLS-F}
where
\be
\epsilon_t(\omega)=\frac{1}{\sqrt{2\pi}}\int_{0}^td\tau
\epsilon(\tau) e^{i\omega \tau}
\ee
is the finite-time Fourier transform of $\epsilon(t)$.

One can consider a more specific scenario, namely, coupling to
zero-temperature ($T=0$) AN bath. The effects of the bath then
amount to the decay of the excited state's population, which can
be written as:
\be
P_e(t)=\exp[-R_e(t)Q(t)]. \e{55}

\section{Modulation arsenal for AN and PN}
\label{Sec-modulation}
Any modulation with quasi-discrete, finite spectrum is deemed
quasiperiodic, implying that it can be expanded as
\be
\label{quasi-def}
\epsilon(t)=\sum_k\epsilon_{k}e^{-i\nu_{k}t}
\ee
where $\nu_{k}\,(k=0,\pm1,...)$ are arbitrary discrete frequencies
such that
\be
\label{quasi-cond}
|\nu_{k}-\nu_{k'}|\geq \Omega \quad \forall k\neq k',
\ee
where $\Omega$ is the minimal spectral interval.

One can define the long-time limit of the quasi-periodic
modulation, when
\be
\label{long-times}
\Omega t\gg 1  \quad{\rm and}\quad t\gg t_c,
\ee
where $t_c$ is the bath-memory (correlation) time, defined as the
inverse of the largest spectral interval over which $G_T(\omega)$
and $G_T(-\omega)$ change appreciably near the relevant
frequencies $\omega_a+\nu_k$. In this limit, the fluence is given
by
\be
\label{eppsilon-c}
Q(t)\approx\epsilon_ct\quad\epsilon_c=\sum_k|\epsilon_k|^2,
\ee
resulting in the average decay rate:
\bea
\label{75}
&&R_e=2\pi\sum_k|\lambda_k|^2G(\omega_a+\nu_k), \\
&&\lambda_k=\epsilon_k/\epsilon_c.
\eea

\subsection{Phase modulation (PM) of the coupling}
\label{Sec-Qubit-PM}

\subsubsection{Monochromatic perturbation}
\label{Sec-Mono}
Let
 \be
 \epsilon(t)=\epsilon_0e^{-i\Delta t}.
 \e{3.1}
 Then
 \be
R_e=2\pi G_T(\omega_a+\Delta),
 \e{3.2}
where $\Delta=\mbox{const.}$ is a frequency shift, induced by the
AC Stark effect (in the case of atoms) or by the Zeeman effect (in
the case of spins).
 In principle, such a shift may drastically enhance or suppress $R$ relative
to the Golden - Rule decay rate, i.e. the decay rate without any
perturbation
\be
R_{\rm GR} = 2\pi G_T(\omega_a). \e{3.2.1}
 Equation \r{3.2} provides the {\em maximal change} of $R$
achievable by an external perturbation, since it does not involve
any averaging (smoothing) of $G(\omega)$ incurred by the width of
$F_t(\omega)$: the modified $R$ can even {\em vanish}, if the
shifted frequency $\omega_a+\Delta$ is beyond the cutoff frequency
of the coupling, where $G(\omega)=0$ (Figure \ref{Fig-1}a). This
would accomplish the goal of dynamical decoupling
\cite{aga99,aga01a,alicki2004oss,vio98,shi04,vit01,fac01,fac04,zan03}.
Conversely, the increase of $R$ due to a shift can be much greater
than that achievable by repeated measurements, i.e. the anti-Zeno
effect \cite{kof00,kof01b,kof01a,kof96}.
 In practice, however, AC Stark shifts are usually small for (cw)
monochromatic perturbations, whence pulsed perturbations should
often be used, resulting in multiple $\nu_k$ shifts, as per
Eq.~\eqref{75}.

\subsubsection{Impulsive phase modulation}
\label{Qubit-Impulsive}

 Let the phase of the modulation function periodically jump by an
amount $\phi$ at times $\tau,2\tau,\dots$.
 Such modulation can be achieved by a train of identical, equidistant,
narrow pulses of nonresonant radiation, which produce pulsed AC
Stark shifts of $\omega_a$.
 Now
\be
 \epsilon(t)=e^{i[t/\tau]\phi},
 \e{3.3}
 where $[\dots]$ is the integer part.
 One then obtains that
  \be
Q(t)=t,\ \ \epsilon_c=1, \e{3.4}
  \be
F_{n\tau}(\omega)=\frac{2\sin^2(\omega\tau/2)
\sin^2[n(\phi+\omega\tau)/2]} {\pi
n\tau\omega^2\sin^2[(\phi+\omega\tau)/2]}. \e{87} The
excited-state decay, according to equation \r{55}, has then the
form (at $t=n\tau$)
\be
P_e(n\tau)=\exp[-R_e(n\tau)n\tau], \e{63} where $R_e(n\tau)$ is
defined by Eqs. \r{53} and \r{87}.

For sufficiently long times (Eq.~\eqref{long-times}) one can use
Eq.~\eqref{75}, with
\be
\nu_k=\frac{2k\pi}{\tau}-\frac{\phi}{\tau},\ \
|\lambda_k|^2=\frac{4\sin^2(\phi/2)}{(2k\pi-\phi)^2} \e{89}

\begin{figure}[htb]
\centering\includegraphics[width=8cm]{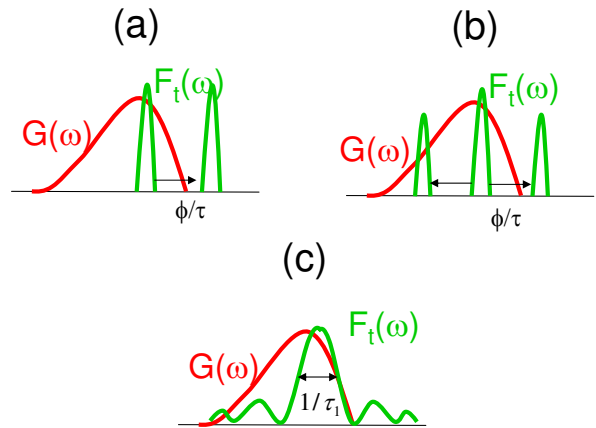}
\protect\caption{Spectral representation of the bath coupling,
$G(\omega)$, and the modulation, $F_t(\omega)$. (a) Monochromatic
modulation, or impulsive phase modulation, with small phase
shifts, $\phi\ll1$, and $1/\tau$ repetition rate. (b) Impulsive
phase modulation, ($\pi$-pulses), $\phi=\pi$. (c) On-off
modulation, with $1/\tau_1$ repetition rate for $\tau_1\ll\tau_0$.
 }\label{Fig-1}\end{figure}

For {\em small phase shifts}, $\phi\ll 1$, the $k=0$ peak
dominates,
\be
|\lambda_0|^2\approx 1-\frac{\phi^2}{12}, \e{3.8} whereas
\be
|\lambda_k|^2\approx\frac{\phi^2}{4\pi^2k^2}\ \ (k\ne 0). \e{3.9}
In this case one can retain only the $k=0$ term in Eq.~\eqref{75},
unless $G(\omega)$ is changing very fast with frequency. Then the
modulation acts as a constant shift, (Fig.~\ref{Fig-1}a)
\be
\Delta=-\phi/\tau. \e{3.10}

As $|\phi|$ increases, the difference between the $k=0$ and $k=1$
peak heights diminishes, {\em vanishing} for $\phi=\pm\pi$. Then
\be
|\lambda_0|^2=|\lambda_1|^2=4/\pi^2, \e{3.11} i.e., $F_t(\omega)$
for $\phi=\pm\pi$ contains {\em two identical peaks symmetrically
shifted in opposite directions} (Figure \ref{Fig-1}b) [the other
peaks $|\lambda_k|^2$ decrease with $k$ as $(2k-1)^{-2}$, totaling
0.19].

The foregoing features allow one to adjust the modulation
parameters for a given scenario to obtain an {\em optimal}
decrease or increase of $R$. Thus, the phase-modulation (PM)
scheme with a small $\phi$ is preferable near a continuum edge
(Figure \ref{Fig-1}a,b), since it yields a spectral shift in the
required direction (positive or negative).
 The adverse effect of $k\ne 0$ peaks in $F_t(\omega)$ then scales as
$\phi^2$ and hence can be significantly reduced by decreasing
$|\phi|$.
 On the other hand, if $\omega_a$ is near a {\em symmetric}
peak of $G(\omega)$, $R$ is reduced more effectively for
$\phi\simeq\pi$, as in Refs. \cite{aga01a,aga01}, since the main
peaks of $F_t(\omega)$ at $\omega_0$ and $\omega_1$ then shift
stronger with $\tau^{-1}$ than the peak at $\omega_0=-\phi/\tau$
for $\phi\ll 1$.

\subsection{Amplitude modulation (AM) of the coupling}
\label{Sec-Qubit-AM}

Amplitude modulation (AM) of the coupling may be applicable to
certain AN or PN scenarios. It arises, e.g., for radiative-decay
modulation due to atomic motion through a high-$Q$ cavity or a
photonic crystal \cite{she92,jap96} or for atomic tunneling in
optical lattices with time-varying lattice acceleration
\cite{fis01,niu98}.
\subsubsection{On-off modulation}
\label{Sec-On-Off}
The simplest form of AM is to let the coupling be turned on and
off periodically, for the time $\tau_1$ and $\tau_0-\tau_1$,
respectively, i.e.,
 \be
 \epsilon(t)=\left\{\begin{array}{ll}
1&\mbox{for}\ n\tau_0<t<n\tau_0+\tau_1,\\
0&\mbox{for}\ n\tau_0+\tau_1<t<(n+1)\tau_0
\end{array}\right.
\e{3.12} ($n=0,1,\dots$). Now $Q(t)$ in \r{55} is the total time
during which the coupling is switched on, whereas
\be
F_{n\tau_0}(\omega)=\frac{2\sin^2(\omega\tau_1/2)
\sin^2(n\omega\tau_0/2)}{\pi
n\tau_1\omega^2\sin^2(\omega\tau_0/2)}, \e{91} so that
\be
P(n\tau_0)=\exp[-R(n\tau_0)n\tau_1], \e{3.21} where $R(n\tau_0)$
is given by Eqs. \r{53} and \r{91}. This case is also covered by
\r{75}, where the parameters are now found to be
 \be
\epsilon_c^2=\frac{\tau_1}{\tau_0},\ \
\nu_k=\frac{2k\pi}{\tau_0},\ \
|\lambda_k|^2=\frac{\tau_1}{\tau_0}\mbox{sinc}^2\left(
\frac{k\pi\tau_1}{\tau_0}\right).
 \e{3.13}

It is instructive to consider the limit wherein $\tau_1\ll\tau_0$
and $\tau_0$ is much greater than the correlation time of the
continuum, i.e., $G(\omega)$ does not change significantly over
the spectral intervals $(2\pi k/\tau_0,2\pi(k+1)/\tau_0)$.
 In this case one can
approximate the sum \r{75} by the integral \r{53} with
 \be
F_t(\omega)\approx(\tau_1/2\pi)\mbox{sinc}^2(\omega\tau_1/2),
\e{3.14}
 characterized by the spectral broadening $\sim 1/\tau_1$
(figure \ref{Fig-1}c).
 Then equation \r{53} for $R$ reduces to that
obtained when ideal projective measurements are performed at
intervals $\tau_1$ \cite{kof00}.
 Thus the AM on-off coupling scheme
{\em can imitate measurement-induced (dephasing) effects} on
quantum dynamics, if the interruption intervals $\tau_0$ {\em
exceed the correlation time of the continuum}.

\section{Multipartite decoherence control}
\label{Sec-multi}
Multipartite decoherence control, for many qubits coupled to
thermal baths, is a much more challenging task than single-qubit
control since: (i) entanglement between the qubits is typically
more vulnerable and more rapidly destroyed by the environment than
single qubit coherence \cite{ban04a,yu04,yu06}; (ii) the
possibility of cross-decoherence, whereby qubits are coupled to
each other through the baths, considerably complicates the
control. We have recently analyzed this situation and extended
\cite{gor06a,gor06b} the decoherence control approach of
Sec.~\ref{Sec-ME}-\ref{Sec-modulation} to multipartite scenarios,
where the qubits are either coupled to zero-temperature baths or
undergoing proper dephasing.

\subsection{Multipartite AN control by off-resonant
modulation: singly excited systems coupled to $T=0$ baths}
\label{Subsec-multi-amplitude}
The decay of a singly excited multi-qubit system (under amplitude
noise) to the ground state, in the presence of off-resonant
modulating fields is described by the following relaxation matrix
\cite{gor06a,gor06b}:
\bea
\label{zero-gen-J-def} &&J_{jj'}(t) = 2\pi
\int_{-\infty}^\infty d\omega
G_{jj'}(\omega)F_{t,jj'}(\omega)\\
&&G_{jj'}(\omega)=\hbar^{-2}\sum_k\mu_{k,j}\mu^*_{k,j'}
\delta(\omega-\omega_k)\\
\label{K-def} &&F_{t,jj'}(\omega)=
\epsilon^*_{t,j}(\omega-\omega_{j})
\epsilon_{t,j'}(\omega-\omega_{j'})
\eea
Here $G_{jj'}(\omega)$ is the coupling spectrum matrix given by
nature and $F_{t,jj'}(\omega)$ is the dynamical modulation matrix,
which we design at will to suppress the decoherence. The diagonal
elements of the decoherence matrix are the time-integrated
individual qubits' decay rates, while the off-diagonal elements
are the cross-relaxation rates, pertaining to the coupling of the
different qubits through the bath: virtual emission into the bath
by qubit $j$ and its virtual reabsorption by qubit $j'$.

As an example, we may control the relaxation matrix elements by
local (qubit-addressing) impulsive phase modulation, (see
Sec.~\ref{Qubit-Impulsive}), described by
\be
\label{epsilon-omega}
\epsilon_{t,j}(\omega)=\frac{\left(e^{i\omega\tau_{j}}-1\right)
\left(e^{i(\phi_{j}+\omega\tau_{j})[t/\tau_{j}]}-1\right)}
{i\omega\left(e^{i(\phi_{j}+\omega\tau_{j})}-1\right)}.
\ee
Here $[...]$ denote the integer part, $\tau_{j}$ and $\phi_{j}$
are the pulse duration and the phase change for particle $j$,
respectively. In the limit of weak pulses, of area
$|\phi_{j}|\ll\pi$, Eq.~\eqref{epsilon-omega} yields
$\epsilon_{t,j}(\omega)\cong\epsilon_{t,j}\delta(\omega-\Delta_{j})$,
where $\Delta_{j}=\phi_{j}/\tau_{j}$ is the effective spectral
shift caused by the pulses.

One can define the fidelity, $F(t)$, total excitation probability,
$F_p(t)$, and the autocorrelation function, $F_c(t)$ as follows:
\bea
&&F(t)=Tr_{\{j\}}\left(\rho(0)\rho(t)\right)\\
&&F_p(t)=Tr_{\{j'\neq j\}}\left({}_j\bra{e}\rho(t)\ket{e}_j\right)\\
&&F_c(t)=F(t)/F_p(t)
\eea
where $Tr_{\{j\}}$ denotes tracing over all qubits. In the absence
of dynamical control, the autocorrelation decays much faster than
the total excitation probability, and is much more sensitive to
the asymmetry between local particle-bath couplings.

Thus, for initial Bell singlet and triplet states, which do not
experience cross-decoherence but only different local decoherence
rates, we find:
\bea
&\ket{\Psi(0)}=1/\sqrt{2}(\ket{g}_A\ket{e}_B\pm\ket{e}_A\ket{g}_B),\\
&F_p(t)=(e^{-2J_A(t)}+e^{-2J_B(t)})/2;\\
&F_c(t)=(1+C(t))/2=1/2+e^{-\Delta J(t)}/(1+e^{-2\Delta J(t)}),\\
&\Delta J(t) = J_A(t)-J_B(t)
\eea
where $C(t)$ is the concurrence \cite{woo98}.

\begin{figure}[ht]
\centering\includegraphics[width=8.5cm]{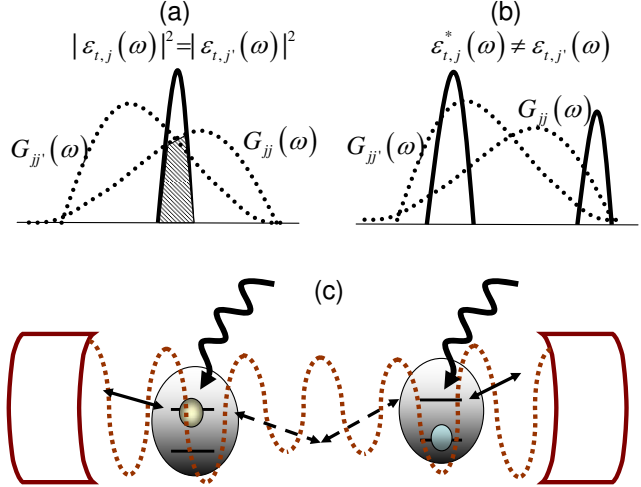}
 \caption{Two two-level particles in a cavity, coupled to the cavity
 modes (thin lines) and subject to local control fields (thick lines). (a,b) Frequency domain
 overlap of coupling spectrum (dotted) and modulation matrix elements(solid),
 resulting in modified decoherence matrix elements (shaded), for:
 (a) global modulation (ICP symmetry), (b) cross-decoherence elimination (IIP symmetry).
 (c) General modulation scheme.}
 \protect\label{Fig-2}
\end{figure}

Without any modulations, decoherence in this scenario has no
inherent symmetry. Our point is that one can symmetrize the
decoherence by appropriate modulations. The key is that different,
``local'', phase-locked modulations applied to the individual
particles, according to Eq.~\eqref{K-def}, can be chosen to cause
{\em controlled interference} and/or spectral shifts between the
particles' couplings to the bath. The $F_{t,jj'}(\omega)$ matrices
(cf.\eqref{K-def}) can then satisfy $2N$ requirements at all times
and be tailored to impose the advantageous symmetries described
below. By contrast, a ``global'' (identical) modulation,
characterized by $F_{t,jj'}(\omega)=|\epsilon_t(\omega)|^2$, is
not guaranteed to satisfy $N\gg1$ symmetrizing requirements at all
times (Fig.~\ref{Fig-2}a).

The most desirable symmetry is that of {\em identically coupled
particles} (ICP), which would emerge if all the modulated
particles could acquire the {\em same} dynamically modified
decoherence and cross-decoherence yielding the following $N\times
N$ fully symmetrized decoherence matrix
\be
\label{J-ICP}
J_{jj'}^{\rm ICP}(t)=r(t)\quad\forall j,j'.
\ee
ICP would then give rise to a $(N-1)$-dimensional decoherence-free
subspace: the entire single-excitation sector less the totally
symmetric entangled state. An initial state in this DFS
\cite{zan97} would neither lose its population nor its initial
correlations (or entanglement).

Unfortunately, it is generally impossible to ensure this symmetry,
since it amounts to satisfying $N(N-1)/2$ conditions using $N$
modulating fields. Even if we accidentally succeed with $N$
particles, the success is not scalable to $N+1$ or more particles.
Moreover, the ability to impose the ICP symmetry by local
modulation fails completely if not all particles are coupled to
all other particles through the bath, i.e. if some
$G_{jj'}(\omega)$ elements vanish.

A more limited symmetry that we may {\em ensure} for $N$ qubits is
that of {\em independent identical particles} (IIP). This symmetry
is formed when spectral shifts and/or interferences imposed by $N$
modulations cause the $N$ different qubits to acquire the {\em
same} single-qubit decoherence $r(t)$ and experience no
cross-decoherence. To this end, we may choose
$\epsilon_{t,j}(\omega)\simeq\epsilon_{t,j}\delta(\omega-\Delta_j)$.
We shall deal with $N$ identical qubits, and set
$\omega_j\equiv\omega_0$. We also require that at any chosen time
$t=T$, the AC Stark shifts satisfy
$\int_0^Td\tau\delta_j(\tau)=2\pi m$, where $m=0,\pm1,...$. This
requirement ensures that modulations only affect the decoherence
matrix \eqref{zero-gen-J-def}, but do not change the relative
phases of the entangled qubits when their MES is probed or
manipulated by logic operations at $t=T$.

\begin{figure}[ht]
\centering\includegraphics[width=8.5cm]{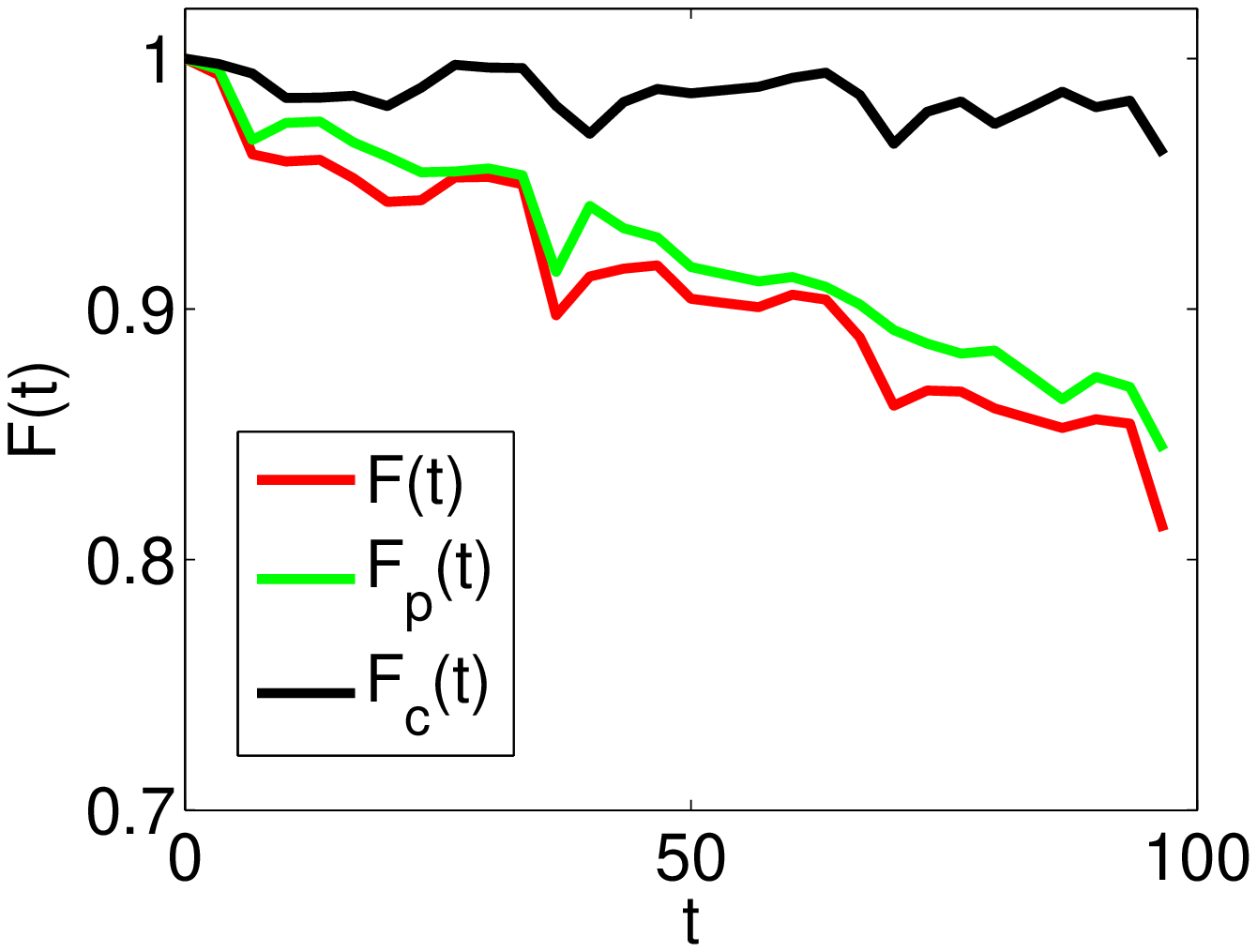}
 \caption{Fidelity of the IIP symmetry for two TLS coupled to zero-temperature baths.
 The initial state is entalged, $\ket{\psi(0)}=1/\sqrt{2}\left(\ket{g}_A\ket{e}_B+\ket{e}_A\ket{g}_B\right)$.}
 \protect\label{Fig-fidelity}
\end{figure}

The spectral shifts $\Delta_j$ can be different enough to couple
each particle to a different spectral range of bath modes so that
their cross-coupling vanishes:
\be
\label{no-cross}
J_{jj'}(t)=2\pi\epsilon^*_{t,j}\epsilon_{t,j'}\int d\omega
G_{jj'}(\omega_0+\omega)\delta(\omega-\Delta_j)\delta(\omega-\Delta_{j'})\rightarrow0.
\ee
Here, the vanishing of $G_{jj'}(\omega)$ for some $j,j'$ is not a
limitation. The $N$ single-particle decoherence rates can be
equated by an appropriate choice of $N$ parameters $\{\Delta_j\}$:
\be
J_{jj'}^{\rm
IIP}(t)=2\pi|\epsilon_{t,j}|^2G_{jj}(\omega_0+\Delta_j)=\delta_{jj'}r(t),
\ee
where $\delta_{jj'}$ is Kronecker's delta (Fig.~\ref{Fig-1}b). The
IIP symmetry results in complete correlation preservation, i.e.
$F_c(t)=1$, but still permits excited-state population loss,
$F_p(t)=e^{-2{\rm Re} \{r(t)\}}$ (Fig.~\ref{Fig-fidelity}). If the
single-particle $r(t)$ may be dynamically suppressed, i.e. if the
spectrally shifted bath response $G_{jj}(\omega_j+\Delta_j)$ is
small enough, this $F_p(t)$ will be kept close to $1$.

\subsection{Multipartite PN control by resonant
modulation}
\label{Subsec-multi-phase}
One can describe phase-noise, or proper dephasing, by a stochastic
fluctuation of the excited-state energy,
$\omega_a\rightarrow\omega_a+\delta_r(t)$, where $\delta_r(t)$ is
a stochastic variable with zero mean, $\mean{\delta(t)}=0$, and
$\mean{\delta(t)\delta(t')}=\Phi^P(t-t')$ is the second moment.
For multipartite systems, where each qubit can undergo different
proper dephasing, $\delta_j(t)$, one has an additional second
moment for the cross-dephasing,
$\mean{\delta_j(t)\delta_{j'}(t')}=\Phi^P_{jj'}(t-t')$. A general
treatment of multipartite systems undergoing this type of proper
dephasing is given in Ref.~\cite{gor06a}. Here we give the main
results for the case of two qubits.

Let us take two TLS, or qubits, which are initially prepared in a
Bell state. We wish to obtain the conditions that will preserve
it. In order to do that, we change to the Bell basis, which is
given by
\bea
\label{dec-two-TLS-Bell-basis}
&&\ket{B_{1,2}}=1/\sqrt{2}e^{i\omega_at}\left(\ket{e}_1\ket{g}_2
\pm
\ket{g}_1\ket{e}_2\right)\\
\label{proper-bell-def}
&&\ket{B_{3,4}}=1/\sqrt{2}\left(e^{i2\omega_at}\ket{e}_1\ket{e}_2
\pm \ket{g}_1\ket{g}_2\right).
\eea
For an initial Bell-state
$\overline{\4\rho}_l(0)=\ket{B_l}\bra{B_l}$, where $l=1...4$, one
can then obtain the fidelity,
$F_l(t)=\bra{B_l}\overline{\4\rho}_l(t)\ket{B_l}$, as:
\bea
\label{proper-two-TLS-F}
&F_{l}(t)=\cos(\phi_\pm(t)){\rm Re}\left[e^{i\phi_\pm(t)}
\left(1-\frac{1}{2}\sum_{jj'}J^P_{jj',l}(t)\right)\right], \eea
where
\bea
&\phi_j(t)=2\int_0^td\tau V_{0,j}(\tau)\\
\label{proper-two-TLS-J-def}
&J^P_{jj',l}(t)=2\pi\int_{-\infty}^\infty d\omega
G^P_{jj'}(\omega)F_{t,jj',l}(\omega)\\
&G^P_{jj'}(\omega)=\int_{-\infty}^\infty dt
\Phi^P_{jj'}(t)e^{i\omega t}\\
\label{proper-two-TLS-Lambda-k-2}
&F_{t,jj,l}(\omega)=|\epsilon_{t,j}(\omega)|^2\\
\label{proper-two-TLS-Lambda-k-3}
&F_{t,jj',3}(\omega)=-F_{t,jj',1}(\omega)=\epsilon^*_{t,j}(\omega)\epsilon^*_{t,j'}(\omega)\\
\label{proper-two-TLS-Lambda-k-4}
&F_{t,jj',4}(\omega)=-F_{t,jj',2}(\omega)=\epsilon_{t,j}(\omega)\epsilon^{*}_{t,j'}(\omega)
\eea
where $V_{0,j}(t)$ is the amplitude of the resonant field applied
on qubit $j$, $\phi_\pm(t)=(\phi_1(t)\pm\phi_2(t))/2$ and the
$\phi_+$ corresponds to $k=1,3$ and $\phi_-$ to $k=2,4$.
Expressions
\eqref{proper-two-TLS-F}-\eqref{proper-two-TLS-Lambda-k-4} provide
our recipe for minimizing the Bell-state fidelity losses. They
hold for {\em any} dephasing time-correlations and {\em arbitrary}
modulation.

\begin{figure}[ht]
\centering\includegraphics[width=8.5cm]{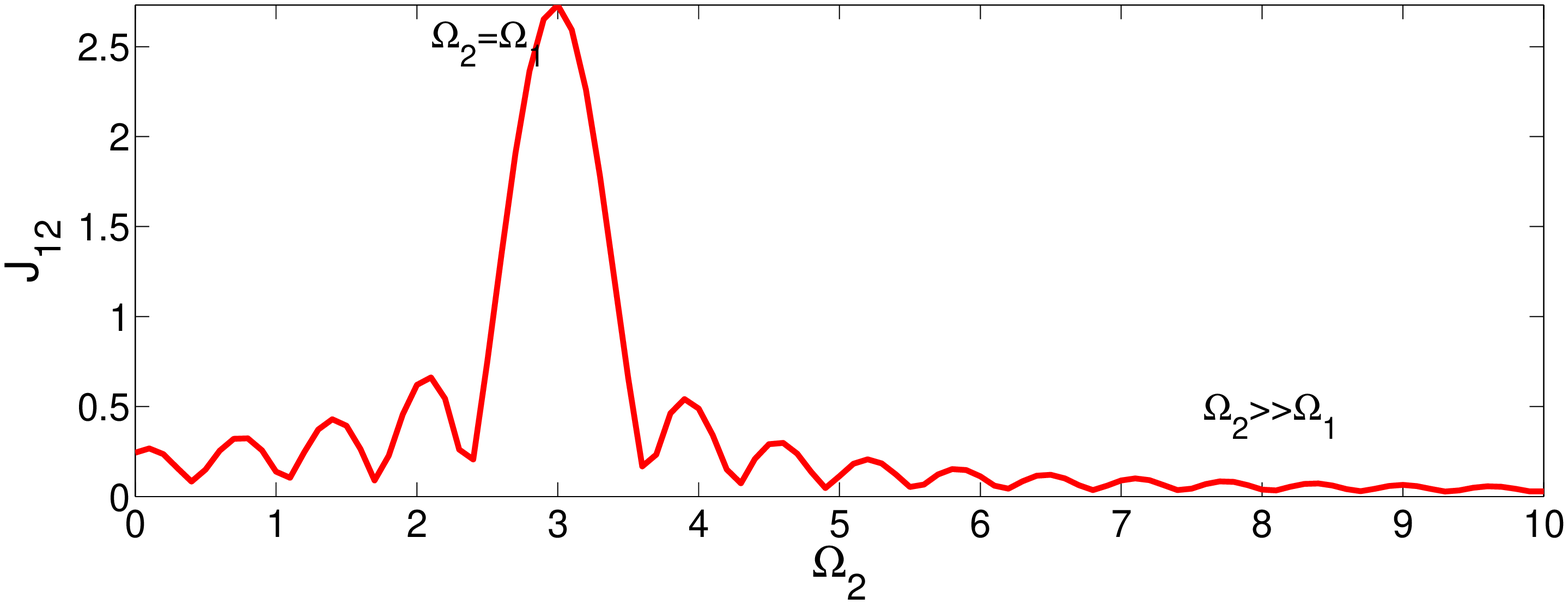}
 \caption{Cross-decoherence as a function of local modulation. Here two qubits are
 modulated by continuous resonant fields, with amplitudes $\Omega_{1,2}$. The cross-decoherence decays
 as the two qubits' modulations become increasingly different. The bath parameters are $\Phi_T(t)=e^{-t/t_c}$, where
 $t_c=0.5$ is the correlation time; and $\Omega_1=3$.}
 \protect\label{Fig-cross}
\end{figure}

One can choose between two modulation schemes, depending on our
goals. When one wishes to preserve and initial quantum state, one
can equate the modified dephasing and cross-dephasing rates of all
qubits, $J_{jj',l}(t)=J(t)$. This results in complete preservation
of the singlet only, i.e. $F_2(t)=1$, for all $t$, but reduces the
fidelity of the triplet state. On the other hand, if one wishes to
equate the fidelity for all initial states, one can eliminate the
cross-dephasing terms, by applying different modulations to each
qubit (Fig.~\ref{Fig-cross}), causing $F_{t,jj',l}(\omega)=0$
$\forall j\neq j'$. This requirement can be important for quantum
communication schemes.

\section{Conclusions}
\label{Sec-conc}
In this paper we have expounded our universal approach to the
dynamical control of qubits subject to AN and PN, by either off-
or on-resonant modulating fields, respectively. It is based on a
general non-Markovian master equation valid for weak system-bath
coupling and arbitrary modulations, since it does not invoke the
rotating wave approximation. The resulting universal convolution
formulae provide intuitive clues as to the optimal tailoring of
modulation and noise spectra. Our analysis of multiple,
field-driven, qubits which are coupled to partly correlated or
independent baths or undergo locally varying random dephasing has
resulted in the universal formula \eqref{zero-gen-J-def} for
coupling to zero-temperature bath, and \eqref{proper-two-TLS-F}
for Bell-state preservation under local proper dephasing.

Our general analysis allows one to come up with an optimal choice
between global and local control, based on the observation that
the maximal suppression of decoherence is not necessarily the best
one. Instead, we demand an optimal {\em phase-relation} between
different, but {\em synchronous} local modulations of each
particle. The merits of local vs. global modulations have been
shown to be essentially twofold:
\begin{itemize}

\item Local modulation can effectively {\em decorrelate} the
different proper dephasings of the multiple TLS, resulting in
equal dephasing rates for all states. For two TLS, we have shown
that the singlet and triplet Bell-states acquire the same
dynamically-modified dephasing rate. This should be beneficial
compared to the standard global ``Bang-Bang'' ($\pi$-phase flips)
if both states are used (intermittently) for information
transmission or storage.

\item For different couplings to a zero-temperature bath, one can
better preserve any initial state by using local modulation which
can reduce the decay as well as the mixing with other states, than
by using global modulation. It was shown that local modulation
which eliminates the cross-decoherence terms, increases the
fidelity more than the global modulation alternative. For two TLS,
it was shown that local modulation better preserves an initial
Bell-state, whether a singlet or a triplet, compared to global
$\pi$-phase ``parity kicks''.

\end{itemize}

{\bf Acknowledgement.} We acknowledge the support of the EC (SCALA
NoE).

\bibliography{Bibliography_JoPB}
\bibliographystyle{unsrt}

\end{document}